\documentclass[a4paper,10pt]{article}

\usepackage{epsf}
\usepackage{graphicx}    

\newcommand{\be}[1]{\begin{equation}\label{#1}}
\newcommand{\ee}{\end{equation}}
\newcommand{\bea}{\begin{eqnarray}}
\newcommand{\eea}{\end{eqnarray}}

\renewcommand{\markright}{\markright{\thepage}}

\begin{document}

\begin{titlepage}

\begin{flushright}
arXiv:0707.4526
\end{flushright}

\vspace{5mm}

\begin{center}

{\Large \bf Statefinder Diagnostic and $w-w^\prime$ Analysis for
 the Agegraphic Dark Energy Models without and with Interaction}

\vspace{7mm}

{\large
 Hao~Wei$^{\,1,}$\footnote{\,email address: haowei@mail.tsinghua.edu.cn}
 and Rong-Gen Cai$^{\,2,}$\footnote{\,email address: cairg@itp.ac.cn}}

\vspace{3mm} {\em $^1$Department of Physics and Tsinghua Center for
 Astrophysics,\\ Tsinghua University, Beijing 100084, China\\
 $^2$Institute of Theoretical Physics, Chinese Academy of Sciences,\\
 P.O. Box 2735, Beijing 100080, China}

\end{center}

\vspace{5mm}
\begin{abstract}
A new dark energy model, named as ``agegraphic dark energy'', has
 been proposed by one of us~(R.~G.~Cai) in arXiv:0707.4049, based
 on the K\'{a}rolyh\'{a}zy uncertainty relation, which arises from
 the quantum mechanics together with general relativity. Then, in
 arXiv:0707.4052, it has been extended by including the interaction
 between the agegraphic dark energy and the pressureless (dark)
 matter. In this note, we investigate the agegraphic dark energy
 models without and with interaction by means of statefinder
 diagnostic and $w-w^\prime$ analysis.\\

 \end{abstract}

 \end{titlepage}

 \newpage

 \setcounter{page}{2}


The cosmological constant problem is
 essentially a problem in quantum gravity, since the cosmological
 constant is commonly considered as the vacuum expectation value
 of some quantum fields. Before a completely successful quantum
 theory of gravity is available, it is more realistic to combine
 the quantum mechanics with general relativity directly.

In general relativity, one can measure the spacetime without
 any limit of accuracy. However, in quantum mechanics, the
 well-known Heisenberg uncertainty relation puts a limit of
 accuracy in these measurements. Following the line of quantum
 fluctuations of spacetime, K\'{a}rolyh\'{a}zy and his
 collaborators~\cite{r1} (see also~\cite{r2}) made an interesting
 observation concerning the distance measurement for Minkowski
 spacetime through a light-clock {\it Gedanken experiment}, namely,
 the distance $t$ in Minkowski spacetime cannot be known to a better
 accuracy than
 \be{eq1}
 \delta t=\lambda t_p^{2/3}t^{1/3},
 \ee
 where $\lambda$ is a dimensionless constant of order unity. We use
 the units $\hbar=c=k_B=1$ throughout this work. Thus, one can use
 the terms like length and time interchangeably, whereas
 $l_p=t_p=1/m_p$ with $l_p$, $t_p$ and $m_p$ being the reduced
 Planck length, time and mass respectively.

The K\'{a}rolyh\'{a}zy relation~(\ref{eq1}) together with the
 time-energy uncertainty relation enables one to estimate a quantum
 energy density of the metric fluctuations of Minkowski
 spacetime~\cite{r3,r2}. Following~\cite{r3,r2}, with respect to
 the Eq.~(\ref{eq1}) a length scale $t$ can be known with a maximum
 precision $\delta t$ determining thereby a minimal detectable cell
 $\delta t^3\sim t_p^2 t$ over a spatial region $t^3$. Such a cell
 represents a minimal detectable unit of spacetime over a given
 length scale $t$. If the age of the Minkowski spacetime is $t$,
 then over a spatial region with linear size $t$ (determining the
 maximal observable patch) there exists a minimal cell $\delta t^3$
 the energy of which due to time-energy uncertainty relation can not
 be smaller than~\cite{r3,r2}
 \be{eq2}
 E_{\delta t^3}\sim t^{-1}.
 \ee
 Therefore, the energy density of metric fluctuations of
 Minkowski spacetime is given by~\cite{r3,r2}
 \be{eq3}
 \rho_q\sim\frac{E_{\delta t^3}}{\delta t^3}\sim
 \frac{1}{t_p^2 t^2}\sim\frac{m_p^2}{t^2}.
 \ee
 We refer to the original papers~\cite{r3,r2} for more details.
 It is worth noting that in fact, the K\'{a}rolyh\'{a}zy
 relation~(\ref{eq1}) and the corresponding energy
 density~(\ref{eq3}) have been independently rediscovered later
 for many times in the literature (see e.g.~\cite{r39,r40,r41}).

In~\cite{r4}, one of us~(R.G.C.) proposed a new dark energy model
 based on the energy density Eq.~(\ref{eq3}). As the most natural
 choice, the length measure $t$ in Eq.~(\ref{eq3}) is chosen to
 be the age of the universe
 \be{eq4}
 T=\int_0^a\frac{da}{Ha},
 \ee
 where $a$ is the scale factor of our universe; $H\equiv\dot{a}/a$
 is the Hubble parameter; a dot denotes the derivative with respect
 to cosmic time. Therefore, we call it as ``agegraphic dark
 energy''. The energy density of the agegraphic dark energy is
 given by~\cite{r4}
 \be{eq5}
 \rho_q=\frac{3n^2m_p^2}{T^2},
 \ee
 where the numerical factor $3n^2$ is introduced to parameterize
 some uncertainties, such as the species of quantum fields in
 the universe, the effect of curved spacetime, and so on. Since
 the energy density~(\ref{eq3}) is derived for Minkowski spacetime,
 the factor $3n^2$ also compiles the effects coming from the
 straightforward application of Eq.~(\ref{eq3}) in the
 Friedmann-Robertson-Walker~(FRW) spacetime. Obviously,
 since the present age of the universe $T_0\sim H_0^{-1}$ (the
 subscript ``0'' indicates the present value of the corresponding
 quantity; we set $a_0=1$), the present energy density of the
 agegraphic dark energy explicitly meets the observed value
 naturally, provided that the numerical factor $n$ is of order
 unity. It is shown that the agegraphic dark energy naturally obeys
 the holographic black hole entropy bound~\cite{r3,r4}, just like
 the holographic dark energy. In addition, by choosing the age of
 the universe rather than the future event horizon as the length
 measure, the drawback concerning causality in the holographic dark
 energy model~\cite{r5} (see also e.g.~\cite{r6,r7} for relevant
 references) does not exist in the agegraphic dark energy
 model~\cite{r4}. The similarity and difference between the
 agegraphic dark energy and holographic dark energy are discussed
 in~\cite{r8}.

We consider a flat Friedmann-Robertson-Walker~(FRW) universe
 containing agegraphic dark energy and pressureless matter, the
 corresponding Friedmann equation reads
 \be{eq6}
 H^2=\frac{1}{3m_p^2}\left(\rho_m+\rho_q\right).
 \ee
 It is convenient to introduce the fractional energy densities
 $\Omega_i\equiv\rho_i/(3m_p^2H^2)$ for $i=m$ and $q$. From
 Eq.~(\ref{eq5}), it is easy to find that
 \be{eq7}
 \Omega_q=\frac{n^2}{H^2T^2},
 \ee
 whereas $\Omega_m=1-\Omega_q$ from Eq.~(\ref{eq6}). By using
 Eqs.~(\ref{eq5}), (\ref{eq7}) and the energy conservation
 equation $\dot{\rho}_m+3H\rho_m=0$, we obtain the equation
 of motion for $\Omega_q$ as~\cite{r4}
 \be{eq8}
 \Omega_q^\prime=\Omega_q\left(1-\Omega_q\right)
 \left(3-\frac{2}{n}\sqrt{\Omega_q}\right),
 \ee
 where a prime denotes the derivative with respect to the
 $e$-folding time $N\equiv\ln a$. From the energy conservation
 equation $\dot{\rho}_q+3H(\rho_q+p_q)=0$, Eqs.~(\ref{eq5})
 and~(\ref{eq7}), it is easy to find that the equation-of-state
 parameter~(EoS) of the agegraphic dark energy
 $w_q\equiv p_q/\rho_q$ is given by~\cite{r4}
 \be{eq9}
 w_q=-1+\frac{2}{3n}\sqrt{\Omega_q}.
 \ee
 In addition, differentiating Eqs.~(\ref{eq6}) and using
 Eqs.~(\ref{eq5}), (\ref{eq7}), and the energy conservation
 equation $\dot{\rho}_m+3H\rho_m=0$, we obtain the deceleration
 parameter as~\cite{r8}
 \be{eq10}
 q\equiv -\frac{\ddot{a}}{aH^2}=-1-\frac{\dot{H}}{H^2}
 =\frac{1}{2}-\frac{3}{2}\Omega_q
 +\frac{\Omega_q^{3/2}}{n}.
 \ee

Notice that the scale factor $a(t)$ can be explicitly expanded as
 $$a(t)=a_0\left[1+H_0(t-t_0)-\frac{1}{2}q_0 H_0^2 (t-t_0)^2+
 \frac{1}{6}\left.\frac{\stackrel{\dots}{a}}{a}\right|_{t_0}
 (t-t_0)^3+\dots\right]$$
 around the present time $t=t_0$. Naturally, the next step
 beyond $H\equiv\dot{a}/a$ and $q\equiv -\ddot{a}/(aH^2)$
 is to consider a new geometrical quantity containing
 $\stackrel{\dots}{a}$. In fact, a so-called statefinder pair
 $\{r,\,s\}$ has been introduced in~\cite{r9}, namely
 \be{eq11}
 r\equiv\frac{\stackrel{\dots}{a}}{aH^3},~~~~~~~
 s\equiv\frac{r-1}{3(q-1/2)},
 \ee
 where $s$ is a combination of $r$ and $q$. So, the coefficient
 of the third term in the Taylor's expansion of $a(t)$ can be
 conveniently expressed as $r_0H_0^3/6$. It is obvious that the
 statefinder is a geometrical diagnostic because it depends only on
 the scale factor $a$. The well-known $\Lambda$CDM model corresponds
 to a fixed point $\{r=1,\,s=0\}$ in the $r-s$ diagram~\cite{r9}.
 Since different cosmological models exhibit qualitatively different
 trajectories of evolution in the $r-s$ plane, the statefinder is
 a good tool to distinguish cosmological models~\cite{r9}. In fact,
 the statefinder diagnostic has been extensively used in many models,
 such as $\Lambda$CDM, quintessence~\cite{r9,r10}, Chaplygin
 gas~\cite{r11,r10}, DGP braneworld~\cite{r9,r10}, interacting
 quintessence model~\cite{r12,r13}, the holographic dark energy
 models without and with interaction~\cite{r14,r15}, the holographic
 dark energy model in non-flat universe~\cite{r16},
 quintom~\cite{r17}, interacting phantom model~\cite{r18},
 five-dimensional cosmology~\cite{r19}, Cardassian model~\cite{r20},
 bulk viscous cosmology~\cite{r21}, tachyon~\cite{r22}, and so on.
 As shown in~\cite{r10,r23}, the statefinder diagnostic combined
 with the future SNAP observation can be used to discriminate
 different cosmological models.


 \begin{center}
 \begin{figure}[htbp]
 \centering
 \includegraphics[width=0.5\textwidth]{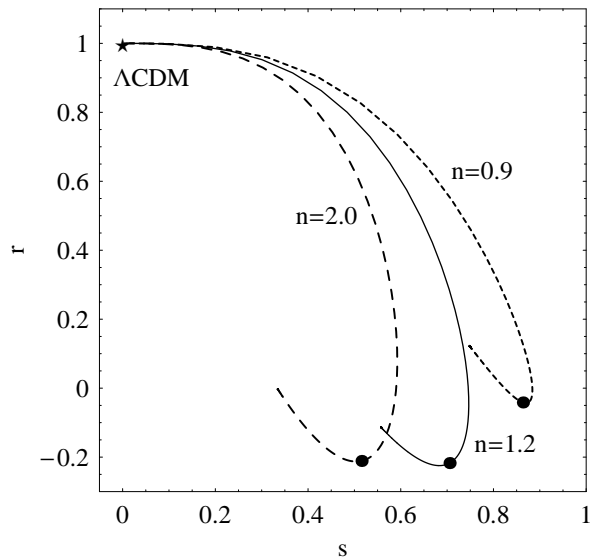}
 \caption{\label{fig1} Evolution trajectories of the statefinder in
 the $r-s$ plane for different model parameter $n$ in
 the agegraphic dark energy model without interaction. The solid
 points indicate the present values of the statefinder. The
 statefiner for the $\Lambda$CDM model is a fixed point and is
 also indicated by a star symbol.}
 \end{figure}
 \end{center}


In this note, we study the agegraphic dark energy by means of
 the statefinder diagnostic. By using the Friedmann equation, i.e.
 Eq.~(\ref{eq6}), and the Raychaudhuri equation
 \be{eq12}
 \dot{H}=-\frac{1}{2m_p^2}\left(\rho_m+\rho_q+p_q\right),
 \ee
 from the definitions Eq.~(\ref{eq11}), one can find
 that~\cite{r9,r10}
 \bea
 &&r=1+\frac{9}{2}\Omega_q w_q (1+w_q)
 -\frac{3}{2}\Omega_q w_q^\prime,\label{eq13}\\
 &&s=1+w_q-\frac{w_q^\prime}{3w_q}.\label{eq14}
 \eea
 Differentiating Eq.~(\ref{eq9}) and using Eq.~(\ref{eq8}), we get
 \be{eq15}
 w_q^\prime=\frac{\sqrt{\Omega_q}}{3n}\left(1-\Omega_q\right)
 \left(3-\frac{2}{n}\sqrt{\Omega_q}\right).
 \ee
 One can solve Eq.~(\ref{eq8}) to get the $\Omega_q(z)$, where $z$
 is the redshift. Note that in the numerical integration of
 Eq.~(\ref{eq8}) we use the initial condition $\Omega_{q0}=0.73$.
 Substituting into Eqs.~(\ref{eq15}), (\ref{eq13}) and~(\ref{eq14}),
 the statefinder $\{r(z),\,s(z)\}$ is in hand. In Fig.~\ref{fig1},
 we show the evolution trajectories of the statefinder in the $r-s$
 plane for different model parameter $n$ in the agegraphic dark
 energy model without interaction. While the universe expands, the
 trajectories of the statefinder start from the point
 $\{r=1,\,s=0\}$ at $z\to\infty$~(i.e. $a\to 0$), then $s$ increases
 to a maximum and $r$ decreases to a minimum, after that the
 trajectories turn a corner and approach another final fixed points
 at $z\to -1$~(i.e. $a\to\infty$). In fact, from Eqs.~(\ref{eq9})
 and~(\ref{eq15}), it is easy to see that $w_q\to -1$ and
 $w_q^\prime\to 0$ while $\Omega_q\to 0$ when $a\to 0$, thus $r\to 1$
 and $s\to 0$ from Eqs.~(\ref{eq13}) and~(\ref{eq14}). The agegraphic
 dark energy mimics the cosmological constant in the early stage.
 Similarly, one can also find out the final fixed point in the $r-s$
 plane. From Eqs.~(\ref{eq9}) and~(\ref{eq15}), we find that
 $w_q\to -1+2/(3n)$ and $w_q^\prime\to 0$ while $\Omega_q\to 1$ when
 $a\to\infty$, therefore $r\to 1+2/n^2-3/n$ and $s\to 2/(3n)$ at that
 time. From Fig.~\ref{fig1}, we see that the trajectories of the
 statefinder for different model parameter $n$ can be significantly
 distinguished. The present value of $s$ is smaller when $n$ is
 larger. It is worth noting that the present values of the
 statefinder for the agegraphic dark energy model without interaction
 are fairly far from the one for the $\Lambda$CDM model, unless $n$
 is very large. Therefore, the statefinder diagnostic combined with
 the future SNAP observation can easily discriminate the agegraphic
 dark energy model without interaction from the $\Lambda$CDM model.


 \begin{center}
 \begin{figure}[htbp]
 \centering
 \includegraphics[width=0.5\textwidth]{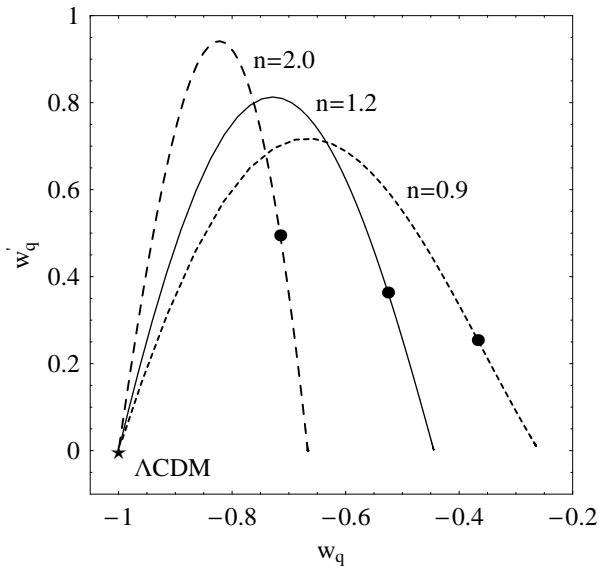}
 \caption{\label{fig2} Evolution trajectories of
 $\{w_q,\,w_q^\prime\}$ for different model parameter $n$ in
 the agegraphic dark energy model without interaction. The solid
 points indicate the present values of $\{w_q,\,w_q^\prime\}$.
 The $\{w,\,w^\prime\}$ of the $\Lambda$CDM model is a fixed
 point and is also indicated by a star symbol.}
 \end{figure}
 \end{center}


In addition to the statefinder $r-s$ which is a geometrical
 diagnostic, there is other dynamical diagnostic $w-w^\prime$ which
 is also used extensively in the literature. Similar to the
 scale factor, the Taylor's expansion of the EoS reads
 $w(N)=w(N=N_0)+w^\prime (N-N_0)/2+\dots$ around $N=N_0$. So, the
 $w-w^\prime$ analysis is important to discriminate models.
 The $w-w^\prime$ analysis was first proposed in~\cite{r24}, and
 then was extended in~\cite{r25,r26}. See
 e.g.~\cite{r27,r28,r29,r30,r31,r32,r15,r16}
 for relevant works, and a recent review can be found in~\cite{r33}.
 The well-known $\Lambda$CDM model corresponds to a fixed point
 $\{w=-1,\,w^\prime=0\}$ in the $w-w^\prime$ plane. In the
 agegraphic dark energy model without interaction, the $w_q$
 and $w_q^\prime$ are given in Eqs.~(\ref{eq9}) and~(\ref{eq15}).
 Again, one can solve Eq.~(\ref{eq8}) to get the $\Omega_q(z)$,
 and then obtain the $w_q(z)$ and $w_q^\prime(z)$. In
 Fig.~\ref{fig2}, we show the evolution trajectories of
 $\{w_q,\,w_q^\prime\}$ for different model parameter $n$ in
 the agegraphic dark energy model without interaction. While
 the universe expands, the trajectories start from the point
 $\{w_q=-1,\,w_q^\prime=0\}$ at $z\to\infty$~(i.e. $a\to 0$),
 then $w_q^\prime$ increases to a maximum, after that the
 trajectories turn a corner and approach another final fixed points
 at $z\to -1$~(i.e. $a\to\infty$). As mentioned above, when
 $a\to 0$, the agegraphic dark energy mimics a cosmological
 constant. From Eqs.~(\ref{eq9}) and~(\ref{eq15}), one can find
 that $w_q\to -1+2/(3n)$ and $w_q^\prime\to 0$ when $a\to\infty$.
 From Fig.~\ref{fig2}, we see that the trajectories of
 $\{w_q,\,w_q^\prime\}$ for different model parameter $n$ can be
 significantly distinguished. The present $w_q$ is smaller and
 the present $w_q^\prime$ is larger when $n$ is larger. In addition,
 we can clearly see that $w_q$ is always larger than $-1$ and cannot
 cross the phantom divide $w=-1$, cf. Eq.~(\ref{eq9}).

In~\cite{r8}, we have extended the original agegraphic dark energy
 model by including the interaction between the agegraphic dark
 energy and the pressureless (dark) matter. We assume that the
 agegraphic dark energy and pressureless (dark) matter exchange
 energy through interaction term $Q$, namely
 \bea
 &&\dot{\rho}_q+3H\left(\rho_q+p_q\right)=-Q,\label{eq16}\\
 &&\dot{\rho}_m+3H\rho_m=Q,\label{eq17}
 \eea
 which preserves the total energy conservation equation
 $\dot{\rho}_{tot}+3H\left(\rho_{tot}+p_{tot}\right)=0$.
 From Eq.~(\ref{eq7}), we get
 \be{eq18}
 \Omega_q^\prime=\Omega_q\left(-2\frac{\dot{H}}{H^2}
 -\frac{2}{n}\sqrt{\Omega_q}\right).
 \ee
 Differentiating Eq.~(\ref{eq6}) and using Eqs.~(\ref{eq17}),
 (\ref{eq5}) and~(\ref{eq7}), it is easy to find that
 \be{eq19}
 -\frac{\dot{H}}{H^2}=\frac{3}{2}\left(1-\Omega_q\right)
 +\frac{\Omega_q^{3/2}}{n}-\frac{Q}{6m_p^2 H^3}.
 \ee
 Therefore, we obtain the the equation of motion for $\Omega_q$ as
 \be{eq20}
 \Omega_q^\prime=\Omega_q\left[\left(1-\Omega_q\right)
 \left(3-\frac{2}{n}\sqrt{\Omega_q}\right)-Q_1\right].
 \ee
 where
 \be{eq21}
 Q_1\equiv\frac{Q}{3m_p^2 H^3}.
 \ee
 If $Q=0$, Eq.~(\ref{eq20}) reduces to Eq.~(\ref{eq8}). From
 Eqs.~(\ref{eq16}), (\ref{eq5}) and~(\ref{eq7}), we get the
 EoS of the agegraphic dark energy as
 \be{eq22}
 w_q=-1+\frac{2}{3n}\sqrt{\Omega_q}-Q_2,
 \ee
 where
 \be{eq23}
 Q_2\equiv\frac{Q}{3H\rho_q}.
 \ee
 Again, if $Q=0$, Eq.~(\ref{eq22}) reduces to Eq.~(\ref{eq9}).
 By using the total EoS $w_{tot}\equiv p_{tot}/\rho_{tot}
 =-1-\frac{2}{3}\frac{\dot{H}}{H^2}=-1/3+2q/3$ and
 $w_{tot}=\Omega_q w_q$, we find that
 \be{eq24}
 q=\frac{1}{2}+\frac{3}{2}\Omega_q w_q.
 \ee

Here, it is also of interest to study the interacting agegraphic
 dark energy model by means of statefinder diagnostic and
 $w-w^\prime$ analysis. From the definition Eq.~(\ref{eq11}), it
 is easy to find that
 \be{eq25}
 r=\frac{\ddot{H}}{H^3}-3q-2.
 \ee
 From Eqs.~(\ref{eq12}), (\ref{eq16}) and~(\ref{eq17}), after some
 algebra, we have
 \be{eq26}
 \frac{\ddot{H}}{H^3}=\frac{9}{2}+\frac{9}{2}\Omega_q w_q (w_q+2)
 -\frac{3}{2}\Omega_q w_q^\prime+\frac{3}{2}Q_1 w_q.
 \ee
 Substituting Eqs.~(\ref{eq24}) and~(\ref{eq26}) into
 Eq.~(\ref{eq25}), we finally obtain that
 \be{eq27}
 r=1+\frac{9}{2}\Omega_q w_q (1+w_q)
 -\frac{3}{2}\Omega_q w_q^\prime+\frac{3}{2}Q_1 w_q.
 \ee
 From the definition Eq.~(\ref{eq11}) and Eqs.~(\ref{eq24}),
 (\ref{eq27}), it is easy to find that
 \be{eq28}
 s=1+w_q-\frac{w_q^\prime}{3w_q}+\frac{Q_1}{3\Omega_q}.
 \ee
 If $Q=0$, Eqs.~(\ref{eq27}) and~(\ref{eq28}) reduce to
 Eqs.~(\ref{eq13}) and~(\ref{eq14}). It is worth noting that the
 forms of $r$ and $s$ in Eqs.~(\ref{eq27}) and~(\ref{eq28}) are
 derived only by using the general Friedmann equation, Raychaudhuri
 equation, and Eqs.~(\ref{eq16}) and~(\ref{eq17}) with a general
 interaction $Q$, and do not depend on any particular dark energy
 model. Therefore, the forms of $r$ and $s$ in Eqs.~(\ref{eq27})
 and~(\ref{eq28}) can be used in {\em any} interacting dark energy
 model, while the subscript ``$q$'' is changed to the corresponding
 ``$de$''. In fact, one can check that the $r$ and $s$ in
 e.g.~\cite{r12,r15} are the special cases of our Eqs.~(\ref{eq27})
 and~(\ref{eq28}) in particular cosmological models with particular
 forms of interaction $Q$.

\begin{center}
 \begin{figure}[htbp]
 \centering
 \includegraphics[width=0.99\textwidth]{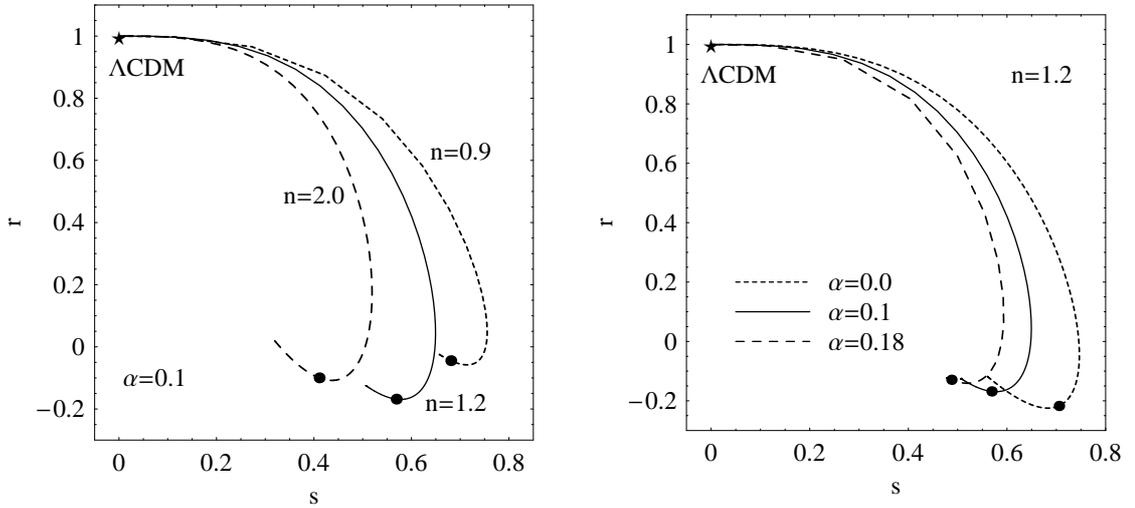}
 \caption{\label{fig3} Evolution trajectories of the statefinder in
 the $r-s$ plane for different model parameters $n$ and $\alpha$ in
 the interacting agegraphic dark energy model. The solid
 points indicate the present values of the statefinder. The
 statefiner for the $\Lambda$CDM model is a fixed point and is
 also indicated by a star symbol.}
 \end{figure}
 \end{center}

In our interacting agegraphic dark energy model, to calculate the
 statefinder, we also need the $w_q^\prime$ in Eqs.~(\ref{eq27})
 and~(\ref{eq28}). From Eqs.~(\ref{eq22}) and~(\ref{eq20}), we have
 \be{eq29}
 w_q^\prime=\frac{\sqrt{\Omega_q}}{3n}\left[\left(1-\Omega_q\right)
 \left(3-\frac{2}{n}\sqrt{\Omega_q}\right)-Q_1\right]-Q_2^\prime.
 \ee
 If the interaction $Q$ is specified, we can solve Eq.~(\ref{eq20})
 to obtain $\Omega_q(z)$. Then, we get the statefinder $\{r,\,s\}$.
 In fact, there are many different forms of interaction $Q$ in the
 literature (see e.g.~\cite{r34,r35,r36,r37,r38} and references
 therein). For convenience, here we only consider a particular
 interaction form $Q=3\alpha H\rho_q$, where $\alpha$ is a
 dimensionless constant. Therefore, in this case,
 \be{eq30}
 Q_1=3\alpha\Omega_q,~~~~~~~Q_2=\alpha,~~~~~~~Q_2^\prime=0.
 \ee
 In principle, $\alpha$ can be positive or negative. However, as
 pointed out in~\cite{r8}, the cases with positive $\alpha$ have
 physically richer phenomena. Thus, in this work, we only consider
 the cases with $\alpha\geq 0$. In Fig.~\ref{fig3}, we show the
 evolution trajectories of the statefinder in the $r-s$ plane for
 different model parameters $n$ and $\alpha$ in the interacting
 agegraphic dark energy model. While the universe expands, the
 trajectories of the statefinder start from the point
 $\{r=1,\,s=0\}$ when $a\to 0$, then $s$ increases to a maximum
 and $r$ decreases to a minimum, after that the trajectories turn
 a corner and approach another final fixed points when
 $a\to\infty$. From Eqs.~(\ref{eq27})---(\ref{eq30}), it is easy
 to find that $r\to 1$ and $s\to 0$ while $\Omega_q\to 0$ when
 $a\to 0$, whereas $r\to 1+2/n^2-3(1+\alpha)/n$ and $s\to 2/(3n)$
 while $\Omega_q\to 1$ when $a\to\infty$. From Fig.~\ref{fig3},
 we see that both $n$ and $\alpha$ significantly affect the
 evolution trajectories of the statefinder in the $r-s$ plane.
 In addition, the present values of the statefinder for the
 interacting agegraphic dark energy model are fairly far from the
 one for the $\Lambda$CDM model, unless $n$ is very large.
 Therefore, the statefinder diagnostic combined with the future
 SNAP observation can easily discriminate the interacting
 agegraphic dark energy model from the $\Lambda$CDM model.


 \begin{center}
 \begin{figure}[htbp]
 \centering
 \includegraphics[width=0.99\textwidth]{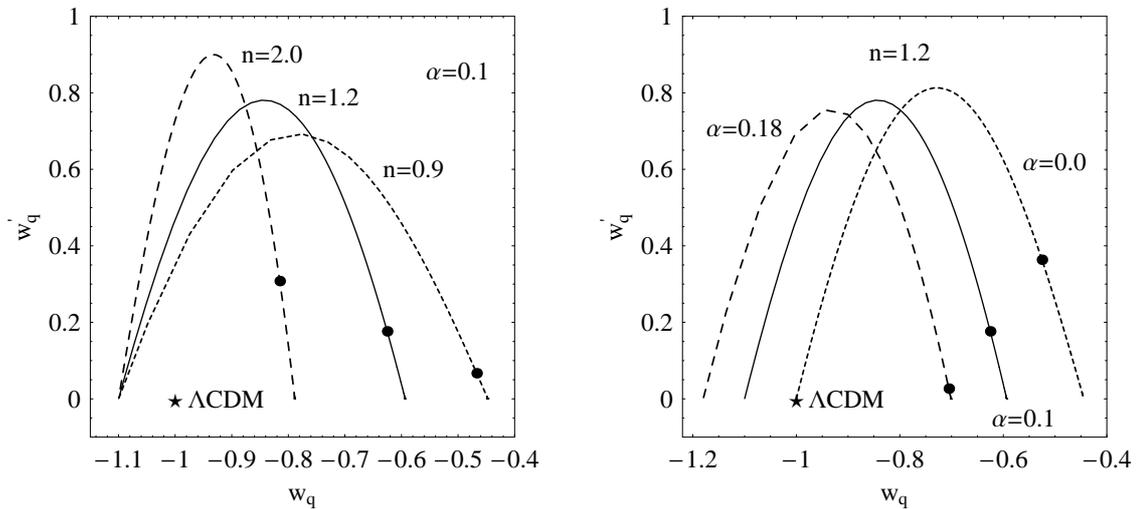}
 \caption{\label{fig4} Evolution trajectories of
 $\{w_q,\,w_q^\prime\}$ for different model parameters $n$ and
 $\alpha$ in the interacting agegraphic dark energy model. The
 solid points indicate the present values of $\{w_q,\,w_q^\prime\}$.
 The $\{w,\,w^\prime\}$ of the $\Lambda$CDM model is a fixed
 point and is also indicated by a star symbol.}
 \end{figure}
 \end{center}


Using Eqs.~(\ref{eq22}) and~(\ref{eq29}), it is easy to obtain the
 evolution trajectories of $\{w_q,\,w_q^\prime\}$ for different
 model parameters $n$ and $\alpha$ in the interacting agegraphic dark
 energy model. We present the results in Fig.~\ref{fig4}. While
 the universe expands, the trajectories start from a fixed point
 when $a\to 0$, then $w_q^\prime$ increases to a maximum, after that
 the trajectories turn a corner and approach another final fixed
 points when $a\to\infty$. From Eqs.~(\ref{eq22}) and~(\ref{eq29}),
 it is easy to find that $w_q\to -1-\alpha$ and $w_q^\prime\to 0$
 while $\Omega_q\to 0$ when $a\to 0$, whereas
 $w_q\to -1-\alpha+2/(3n)$ and $w_q^\prime\to 0$ while $\Omega_q\to 1$
 when $a\to\infty$. Again, from Fig.~\ref{fig4}, we see that both $n$
 and $\alpha$ significantly affect the evolution trajectories of
 $\{w_q,\,w_q^\prime\}$ in the $w_q-w_q^\prime$ plane. The most
 important observation from Fig.~\ref{fig4} is that $w_q$ crossed
 the phantom divide $w=-1$ for the cases with $\alpha\not=0$, which
 is impossible in the agegraphic dark energy model without
 interaction~\cite{r8} (cf. Fig.~\ref{fig2}). The interaction
 significantly changes the situation.

In summary, we investigated the agegraphic dark energy models
 without and with interaction by means of statefinder diagnostic and
 $w-w^\prime$ analysis in this work. Both the statefinder $\{r,\,s\}$
 and the $\{w_q,\,w_q^\prime\}$ can be extracted from some future
 astronomical observations, especially the SNAP-type
 experiments~\cite{r10,r23}. Therefore, they can be used to
 discriminate different cosmological models. In this work, our
 results suggest that the future SNAP observation can easily
 discriminate the agegraphic dark energy model from the
 $\Lambda$CDM model. In addition, since both the model parameters
 $n$ and $\alpha$ significantly affect the evolution trajectories in
 the $r-s$ and $w_q-w_q^\prime$ planes, the future astronomical
 observations can also discriminate the agegraphic dark energy
 models with different parameters.

After all, we admit that although the statefinder diagnostic and
 $w-w^\prime$ analysis are commonly believed to be useful tools
 to discriminate different cosmological models~\cite{r10,r23,r24},
 there are still some different attitudes~(see~\cite{r42} for
 instance). We hope that the future cosmological observations
 which are more precise can shed new light on this issue.


\section*{Acknowledgments}
We are grateful to Prof. Shuang~Nan~Zhang for helpful discussions.
 We also thank Hui~Li, Yi~Zhang, Xing~Wu, Bin~Hu, Xin~Zhang and
 Jingfei~Zhang for useful discussions. This work was supported in
 part by a grant from China Postdoctoral Science Foundation, a grant
 from Chinese Academy of Sciences~(No.~KJCX3-SYW-N2), and by NSFC
 under grants No.~10325525, No.~10525060 and No.~90403029.


\end{document}